\documentclass[12pt,preprint]{aastex}
\usepackage{graphicx}
\usepackage{amsmath}
\usepackage{amsfonts}
\usepackage{amssymb}%
\providecommand{\U}[1]{\protect\rule{.1in}{.1in}}

\newcommand{\beq}{\begin{equation}}
\newcommand{\eeq}{\end{equation}}
\newcommand{\ba}{\begin{array}}
\newcommand{\ea}{\end{array}}

\newcommand{\ee}{\epsilon_{e,0}}

\def\be{\begin{equation}}
\def\ee{\end{equation}}

\def\gsim{\raisebox{-0.3ex}{\mbox{$\stackrel{>}{_\sim} \,$}}}

\begin{document}

\title{Poynting flux dominated jets challenged by their photospheric emission}
\shorttitle{Photospheric emission of Poynting dominated outflows}
\author{D. B\'egu\'e \altaffilmark{1,2,3}, A. Pe'er \altaffilmark{4}}
\shortauthors{B\'egu\'e \& Pe'er}

\altaffiltext{1}{University of Roma ``Sapienza'', 00185, p.le A. Moro 5, Rome, Italy}
\altaffiltext{2}{ICRANet, 65122, p.le della Repubblica, 10, Pescara Italy}
\altaffiltext{3}{Erasmus Mundus Joint Doctorate IRAP PhD student}
\altaffiltext{4}{Physics Department, University College Cork, Cork, Ireland}

\begin{abstract}

One of the key open question in the study of jets in general, and jets
in gamma-ray bursts (GRBs) in particular, is the magnetization of the
outflow.  Here we consider the photospheric emission of Poynting flux
dominated outflows, when the dynamics is mediated by magnetic
reconnection.  We show that thermal three-particle processes,
responsible for the thermalization of the plasma, become inefficient
at a radius $r_{\rm sup} \sim 10^{9.5}$~cm, far below the photosphere,
at $\sim 10^{11.5}$~cm. Conservation of the total photon number above
$r_{\rm sup}$ combined with Compton scattering below the photosphere
enforces kinetic equilibrium between electrons and photons. This, in
turn, leads to an increase in the observed photon temperature, which
reaches $\gtrsim 8$~MeV (observed energy) when decoupling the plasma
at the photosphere. This result is weakly dependent on the free model
parameters. We show that in this case, the expected thermal luminosity
is a few \% of the total luminosity, and could therefore be detected.
The predicted peak energy is more than an order of magnitude higher
than the observed peak energy of most GRBs, which puts strong
constraints on the magnetization of these outflows.

\end{abstract}
\maketitle

\section{Introduction}

One of the key open questions in the study of relativistic outflows in
general, and of gamma-ray bursts (GRBs) in particular, is the
mechanism responsible for accelerating the plasma to the
ultra-relativistic speeds observed, $\Gamma \gsim 100$ \citep[for a
  review, see, e.g.,][]{Meszaros06, GM12, Zhang14}.  In the classical GRB
``fireball'' model \citep{pac86, pac90, RM92, PSN93, RM94}, the
outflow is accelerated by the radiative pressure of the photons
produced during the initial phase of collapse and explosion. In this
model, conservation of energy and entropy implies a linear increase of
the jet Lorentz factor $\Gamma$ with radius $r$ until the jet reaches
the saturation radius $r_s$, above which the jet internal energy is
comparable to its kinetic energy, and no further acceleration is
possible. In this model, magnetic fields are sub-dominant (the energy
density stored in the magnetic field is much smaller than the energy
density in the thermal photon field, $u_{B} \ll u_{th}$).

On the other hand, it was proposed that GRB outflows be magnetically
dominated, $U_{B} \gg U_{th}$ \citep{SDD01, Drenkhahn02, DS02, LB03,
  Giannios06}. In this scenario, if the central engine is able to
produce a highly variable magnetic field, magnetic reconnection may be
the mechanism responsible for the jet acceleration. Under the
assumption of steady energy transfer rate, the most efficient
configuration of magnetic lines orientation leads to a slower increase
in the bulk Lorentz factor with radius, $\Gamma \propto r^{1/3}$ below
the saturation radius \citep{DS02, Drenkhahn02, MR11}.

Since the magnetic field is not directly observed, one has to deduce
its significance indirectly. For example, in their analysis of
GRB080916C, \citet{ZP09} argued that at least part of the outflow
energy has to be in magnetic form. Their argument is based on the
absence of a thermal component in the spectrum, which originates from
the photosphere, and must accompany any photon-dominated outflow.

A different result was recently claimed by \citet{Bromberg+14}. They
argued that highly magnetized jets are disfavored by many GRB
observations, since they do not allow to reproduce the plateau in the
distribution of the GRB duration. This is because Poynting flux
dominated jets are stable and break the envelope of their progenitor
star on a time that is significantly shorter than observed. On
contrast, the break time of baryonic jets are in agreement with the
duration of the observed plateau, favoring this last model.

Solving this controversy is indeed of high importance, as the
magnetization of the outflow puts strong constraints not only on
possible acceleration mechanisms, but also on the nature of GRB
progenitors, as well as the central engines that power GRBs.  In this
paper, we propose a novel way of constraining the magnetization
of GRB outflows, based on their observed spectra. The key is the study
of photon production processes.  As we show here, models in which GRB
jets are strongly magnetized lead to suppression of photon
production. The produced photons, in turn, are Compton up-scattered;
due to their small number, the predicted spectral peak is at $\gsim
8$~MeV, more than an order of magnitude above the typical observed peak.

\section{Dynamics of Poynting-flux dominated jets}
\label{sec:model}

The evolution of the hydrodynamic quantities in a Poynting flux
dominated outflow was first derived by \citet{Drenkhahn02, DS02}, and
was further discussed by \citet{Giannios05, Giannios06, GS05, MR11}.  In
this model, an important physical quantity is the magnetization
parameter, $\sigma_0$, which is the ratio of Poynting flux to kinetic
energy flux at the Alfv\'en point, $r_0$ \footnote{In Poynting flux
  dominated models, at $r_0$ the flow velocity is equal to the
  Alfv\'en speed. Acceleration takes place at $r_0 < r < r_s$, where
  $r_s$ is the saturation radius.}. This quantity plays a similar role
to that of the baryon loading, in the classical ``fireball'' model.

The magnetic field in the flow changes polarity on a small scale,
$\lambda$, which is of the order of the light cylinder in the central
engine frame ($\lambda \approx 2 \pi c / \Omega$, where $\Omega$ is
the angular frequency of the central engine \citep[presumably a
  spinning black hole; see][]{Coroniti90}). This polarity change leads
to magnetic energy dissipation via reconnection process, that is
modeled by a fraction $\epsilon$ of the Alfv\'en speed \footnote{Note
  that this prescription assumes constant rate of energy transfer
  along the jet. As the details of the reconnection process are
  uncertain, the value of $\epsilon$ is highly uncertain. Often a
  constant value $\epsilon \approx 0.1$ is assumed in the
  literature. Further, note that the Alfv\'en speed is essentially
  equal to the speed of light in magnetically dominated outflows.}.

The dissipated magnetic energy is converted to kinetic energy of the
outflow, leading to an acceleration of the plasma. The spatial
evolution of the Lorentz factor, $\Gamma$ and of the comoving number
density, $n_e'$, below the saturation radius $r_s$ are given by
\begin{align}
\Gamma(r) & = \Gamma_\infty \left ( \frac{r}{r_s}\right
)^{1/3} \text{,}
\label{eq:1}\\
n_e' & = \frac{L}{m_p c^3 r^2 \Gamma(r) (\sigma_0+1)^{3/2} }
\text{,}
\label{eq:2}
\end{align}
where the terminal Lorentz factor $\Gamma_{\infty} \simeq
\sigma_0^{3/2}$. The saturation radius is given by $r_s = \pi c
\Gamma_\infty^2 /(3\epsilon\Omega)$, where $\epsilon \Omega$ is the
characteristic frequency of the reconnection
process \citep{Drenkhahn02}. The outflow luminosity (both magnetic
and kinetic per unit of solid angle $\text{str}^{-1}$) is $L = L_k + L_B = \dot{M} c^2 (\sigma_0 +
1)^{3/2}$, where $\dot{M}$ is the outflow mass flux.

The spatial evolution of the comoving magnetic field, $B' = B/\Gamma$
can be calculated using Equation \ref{eq:1}, the definition of the
magnetic luminosity, $L_B = (B^2 / 4 \pi) c r^2$, the definition of
the saturation radius, and energy conservation $L=L_{k}+L_B$. Using
$L_B(r) = L(1-\Gamma(r)/\Gamma_\infty) = L(1- \Gamma(r) /
\sigma_0^{3/2})$ below the saturation radius \citep{GS05}, one obtains
\begin{align}
B' \equiv {B \over \Gamma} & = \left( {4 \pi L \over c} \right)^{1/2}
\left({ \pi c \over 3} \right)^{1/3} {1 \over r^{4/3} \sigma_0^{1/2}
  (\epsilon \Omega)^{1/3}} \left[ 1-\left( \frac{r}{r_s}\right )^{1/3} \right]^{1/2} \nonumber \\
  & \approx 1.4 \times 10^8 \frac{L_{52}^{1/2}}{
  r_{11}^{4/3 } (\epsilon \Omega)_3^{1/3} \sigma_2^{1/2}}~{\rm G}
\text{,} 
\label{Eq:comovingB} 
\end{align}
where $r_s \gg r$ is taken in the last equality, and $Q = 10^x Q_X$ in
cgs units is used here and below.

Deep enough in the flow, radiation and matter are in thermodynamic
equilibrium, sharing the same temperature, $T$. The thermal energy
increases by magnetic energy dissipation, and simultaneously decreases
due to adiabatic losses. As a consequence, only a fraction of the
injected thermal energy appears as black body radiation at the
photospheric radius where matter and radiation decouple.

The spatial evolution of the comoving temperature was calculated by
\citet{GS05}, under the assumption of full thermalization. For
completeness, we briefly repeat their arguments. 
For constant magnetic energy dissipation rate, the energy released at
radii ($r.. r+ dr$) is 
\begin{align}
d \dot{E} = \left ( -\frac{dL_B}{dr} \right ) dr = \frac{L}{3 \sigma_0} \left ( \frac{3}{\pi c}\right )^{1/3} (\epsilon \Omega)^{1/3} r^{-2/3} dr,
\label{eq:ddotE}
\end{align}
where we used the formula for $L_B$  given above Equation \ref{Eq:comovingB},
Equation \ref{eq:1} and the definition of $r_s$. About half of this
dissipated energy is used to accelerate the flow, and the other half
increases its thermal energy \citep{SD04}. Adiabatic losses in
radiative dominated flow imply $T' \propto {n_e'}^{1/3} \propto
r^{-7/9}$, using Equation \ref{eq:2}. Using again the scaling of the
Lorentz factor in Equation \ref{eq:1}, one obtains $L_{th}(r) \propto
r^2 \Gamma^2 {T'}^4 \propto r^{-4/9}$. Therefore, by the time the
plasma reaches some radius $R$, only a fraction $(r/R)^{4/9}$ of the
energy dissipated at radius $r < R$ is still in thermal form.  Integrating over
all radii, the thermal luminosity at radius $r$ is given by 
\beq
L_{th}(r) = {1 \over 2} \int_{r_0}^r d \dot{E} \left({r' \over r}
\right)^{4/9} dr' = {1 \over 2} \left({L \over 3 \sigma_0}\right)
\left({3 \over \pi c} \right)^{1/3} (\epsilon \Omega)^{1/3} \left({9
  \over 7}\right) r^{1/3}.
\label{eq:L_ph}
\eeq
The comoving temperature of the flow is calculated using $L_{th} = {16
  \over 3} \sigma_{SB} r^2 \Gamma^2 {T^{'}}^4$, where $\sigma_{SB}$ is
Stefan-Boltzmann constant, and is given by
\beq
\theta' \equiv {k_B T' \over m_e c^2} = 1.4 \times 10^{-3}
\frac{L_{52}^{1/4}}{r_{11}^{7/12} (\epsilon \Omega)_3^{1/12}
  \sigma_2^{1/2} } \text{,}
\label{eq:theta}
\eeq
where we normalized the temperature to natural units of $m_e c^2$.

Photons decouple the plasma once they reach the photosphere, at which
the optical depth becomes smaller than the unity. Along the radial
direction, $d\tau = \Gamma(1-\beta) n_e' \sigma_T dr$, where $\sigma_T$
is Thomson's cross section. Integrating from $r_{ph}$ to infinity and
requiring $\tau(r_{ph}) = 1$, using Equations \ref{eq:1} and
\ref{eq:2}, the photospheric radius is given by \citep{ANP91, GS05,
  Peer08}
\beq
r_{ph} = 6 \times 10^{11} \frac{L_{52}^{3/5}}{(\epsilon
  \Omega)_3^{2/5} \sigma_2^{3/2}}~\rm{cm}.
\label{eq:r_ph}
\eeq
For the fiducial values of the free model parameters assumed,
$\sigma_0 = 100$ and $(\epsilon \Omega) = 10^3$, this radius is below
the saturation radius, $r_s \approx 10^{13.5}~\sigma_2^3 {(\epsilon
  \Omega)}_3^{-1}$~cm. This implies that the photons decouple the
  plasma while it is still in the acceleration phase.

The results of Equation \ref{eq:theta} imply that as long as the
photons maintain thermal equilibrium, their comoving number density
scales with radius as $n'_\gamma \propto u_{th}'/\langle \epsilon'
\rangle \propto {\theta'}^4/\theta' \propto r^{-7/4}$. Here, $u_{th}'
= a T'^4$ is the comoving thermal energy density, and $ \langle
\epsilon' \rangle = 2.7 k_B T'$ is the average photon energy. If,
however, photon production is suppressed above some radius $r_{\rm sup} <
r_{ph}$ (namely, the remaining photons are still coupled to the
particles in the plasma), the scaling low $T' \propto r^{-7/9}$
derived above implies that the photon density changes with radius as
$n'_\gamma \propto T'^3 \propto r^{-7/3}$. The photon number density
in this case thus drops faster than in thermal equilibrium. These
photons eventually decouple the plasma at the photosphere. As we show
below, this different scaling law modifies the emerging spectra at the
photosphere, and in particular the observed peak energy \footnote{We
  note that in the classical ``fireball'' model dynamics, where
  magnetic fields are sub-dominant, this does not hold: even if photon
  production is suppressed above a certain radius, the scaling laws of
  the photon number density below the photosphere is not affected.}.

\section{Photon production mechanisms}
\label{sec:photon_production}

In the following, we consider photon production below the
photosphere. The leading radiative processes are double Compton,
bremsstrahlung, and cyclo-synchrotron.  Other radiative mechanisms,
such as radiative pair production and three-photon annihilation are
discarded because the plasma is not relativistic ($\theta' < 1$, see
Equation \ref{eq:theta}).

The key question is whether the photon sources are capable of
producing enough photons to enable full thermalization below the
photosphere.  The rate of the interactions considered below were
discussed by \citet{Beloborodov13} and \citet{VLP13}, and references
  therein. For each of these processes, the radius at which a given
interaction freezes out is given by equating the photon production
rate $\dot{n}$ to the expansion rate,
\begin{align}
t_{\rm exp}\dot{n} \geq n_{\gamma, th} \label{Eq:Eqrate}
\end{align}
where $t_{\rm exp} = r/(c\Gamma(r))$ and $n_{\gamma, th} = 16 \pi
\zeta(3) (k_B T')^3 / (c h)^3$ is the photon number density obtained
if the photons are in thermal equilibrium ($\zeta(3) \approx 1.202$ is
the Riemann zeta function and $h$ is Planck's constant).

\emph{Double Compton.} The rate of photon production in double Compton
process is given by \citep{Lightman81}
\begin{align}
\dot n_{DC} = \frac{16 \alpha}{\pi} c \sigma_T {\theta^{'}}^2
g_{DC}(\theta^{'}) \ln \left ( \frac{k_B T^{'}}{E_0} \right ) n_e'
n_{\gamma,th},
\label{Eq:rateDC}
\end{align}
where $\alpha$ is the fine-structure constant, $g_{DC}(\theta^{'}) =
(1+13.91 \theta^{'} + 11.05{\theta^{'}}^2+19.92 {\theta^{'}}^3)^{-1}
\approx 1$ is a fitted formula to the exact numerical result
\citep{Svensson84} and $E_0$ is the threshold energy\footnote{A
  photon of energy $E > E_0$ will be up-scattered to higher energy by
  single Compton scattering and avoid re-absorption by the inverse
  process. $E_0$ is found by equating the Compton parameter $y=4
\theta^{'} c \sigma_T n^{'}$ to the photon opacity. For the double
Compton process, $E_0$ is such that $(E_0 / k_B T^{'})^2 = 9.6 \alpha
\theta^{'} g_{DC}(\theta^{'}) / \pi$.}. 
Using Equations \ref{Eq:Eqrate} and \ref{Eq:rateDC}, the radius
$R_{DC}$ at which double Compton freezes out is
\begin{align}
R_{DC} = 2.4 \times 10^9~L_{52}^{\frac{9}{17}}\, \sigma_2^{-\frac{21}{17}}\, (\epsilon\Omega)^{-\frac{5}{17}}~\text{cm.}
\end{align}

\emph{Bremsstrahlung.}  The temperature at which Bremsstrahlung
freezes out is not relativistic, hence the pair density is expected to
be much smaller than the proton density. As a consequence, the
dominant bremsstrahlung process is scattering between electrons and
protons. The rate of photon production via $e-p$ bremsstrahlung can be
derived, e.g., using formula (5.14) in \citep{RL79}. Dividing by $h
\nu$ and using the normalized photon energy $x \equiv h \nu/ m_e c^2$,
the photon emission rate per unit volume per unit energy is $d {\dot
  n} /dx = (8/3\pi)^{1/2} c \sigma_T \alpha {n_e'}^2 \theta'^{-1/2}
x^{-1} \bar{g}_{ff}$, where the Gaunt factor can be approximated by
$\bar{g}_{ff} \simeq (\sqrt{3}/\pi) \ln (2.25 \theta/x)$
\citep{NT73, IS75,PSS83}. The total photon emission rate is
calculated by integrating over all energies, from $x=x_{\min}$ to
$x=\theta$,
\begin{align}
\dot{n}_{B} = {\sqrt{2} \over \pi^{3/2}} c \sigma_T \alpha
\theta'^{-1/2} {n_e'}^2  \left[ \ln \left( 2.25 \frac{\theta'}{x_{\min}}
  \right)^2 - \ln (2.25)^2 \right]
\label{eq:rate_nb}
\end{align}
The lower boundary on the energy of emitted photons, $x_{\min}$ is
found by comparing the absorption time, $(\alpha_{ff} c)^{-1}$ to the
typical time a photon gains sufficient energy (by inverse Compton
scattering) to avoid re-absorption, $4 \theta' n c \sigma_T$
\citep{VLP13}. Using standard formula for free free absorption in the
Rayleigh-Jeans limit, $x_{\min}$ is calculated by solving
\begin{align}
x_{\min}^2 = \frac{1}{8\sqrt{2} \pi^{5/2}} \alpha \lambda_c^3 n^{'}
{\theta^{'}}^{-5/2} \ln \left (2.25 \frac{\theta'}{x_{\min}}\right),
\label{eq:x_min}
\end{align}
where $\lambda_c = h/m_e c$ is the Compton wavelength.  While an
analytic solution to this Equation does not exist, it is easily
checked numerically that for a wide range of relevant parameter
space, $10^{-4} \lesssim x_{\min} \lesssim 10^{-3}$, leading to $\bar A
  \equiv \ln (2.25 \theta'/x_{\min})^2 - \ln (2.25)^2 \simeq 15$.

Using these results in Equation \ref{Eq:Eqrate} enables to calculate
the radius at which bremsstrahlung freezes out. For $\sigma_0 \gg 1$,
this radius is approximated by
\begin{align}
R_B \simeq 2.47 \times 10^9 \left (\frac{\bar{A}}{15} \right )^{\frac{24}{47}}\,
  L_{52}^{\frac{27}{47}} (\epsilon \Omega)_3^{\frac{7}{47}} \sigma_2^{-\frac{30}{47}} \text{ ~ cm.}
\label{eq:r_B}
\end{align}

\emph{Cyclo-synchrotron.} The rate of photon emission via
cyclo-synchrotron process from a thermal population of electrons is
given by \citep[][and references therein]{VLP13} 
\begin{align}
\dot{n}_{CS} = \frac{12 \pi m_e}{h^3} \sigma_T n_e' \theta'^2 \hat{E_0}^2,
\end{align}
where $\hat{E_0}$ is the energy at which up-scattering and re-absorption
rates are equal. For $\theta' \ll 1$,  $\hat{E_0}$ can be approximated by \citep{VLP13}
\begin{align}
\hat{E_0} &= 14 \left( \frac{m_e c^2}{E_B} \right)^{1/10} \theta'^{3/10} E_B \text{,}
\end{align}
where $E_B = h q B'/(2 \pi m_e c)$ is the cyclotron energy in the
comoving frame.  Assuming $\sigma_0 \gg 1$, and using the equations
above, one finds that the freeze-out radius for cyclo-synchrotron
emission is
\begin{align}
R_{CS} = 5.50 \times 10^{9} L_{52}^{\frac{54}{115}} (\epsilon \Omega)_3^{-\frac{37}{115}} \sigma_2^{-\frac{96}{115}} \text{cm.}
\label{eq:R_cs}
\end{align}

While in the derivation of equation \ref{eq:R_cs} we assumed a thermal
population of electrons, we do not expect this result to change if
electrons are accelerated to high energies during the dissipation
process. This is due to the fact that the typical energy of a
synchrotron emitted photon is proportional to $\gamma_{el}^2$, where
$\gamma_{el}$ is the Lorentz factor associated with the random motion
of the electrons, and the total radiated power is, similarly, proportional to $
\gamma_{el}^2$. Thus, the rate of photon emission is independent on
$\gamma_{el}$.

All radiative process freeze out at $r_{\rm sup}=\max(R_{CD}, R_B,
R_{CS})$. For the fiducial values of the luminosity, magnetization and
angular frequency, Equations \ref{eq:r_ph}, \ref{Eq:rateDC},
\ref{eq:r_B} and \ref{eq:R_cs} imply $r_{\rm sup} \ll r_{ph}$. As a
result, thermal equilibrium can exist only at radii $r \leq
r_{\rm sup}$. Above this radius, photons are not emitted at a high enough
rate to ensure full thermalization. However, below the photosphere,
Compton scattering enforces kinetic equilibrium between electrons and
photons, such that both components can be described by a single
temperature. The photon distribution at $r_{\rm sup} < r < r_{ph}$
therefore obeys a Wien statistics.

\section{Consequences of photon starvation}

The results of the previous section imply that to a good
approximation, one can assume that at $r > r_{\rm sup}$ the total
number of photons is conserved. The photons thus follow a Wien
distribution, with average (co-moving) photon energy $\langle
\epsilon' \rangle \sim 3 k_B T'$. Due to the strong coupling between
photons and electrons below the photosphere, the comoving thermal
energy $u_{th}'$ is shared by the protons, electrons and photons. As
the plasma is non-relativistic, the energy density at $r=r_{\rm sup}$
is $u_{th}'(r_{\rm sup}) = 3 k_B T'_{\rm sup} (n_e' + n_{\gamma})$,
where $T'_{\rm sup} = T'(r=r_{\rm sup})$, and the electron and photon
densities are evaluated at $r_{\rm sup}$.

At larger radii, $r \geq r_{\rm sup}$, full thermalization cannot be
achieved.  Nonetheless, due to the strong coupling between electrons
and photons, for radii not much above $r_{\rm sup}$ (see below) the
photon distribution is close to thermal, with comoving temperature
given by $k_B T'(r) = u'_{th}(r) / 3 ( n_e'(r) + n_\gamma(r) )$. The
energy density is $u_{th}'(r) = L_{th}(r)/ (4/3) r^2 \Gamma^2(r)
c$, \footnote{We omit the factor $\pi$ in the denominator, as
  $L_{th}(r)$ is the luminosity per steradian.} with $L_{th}(r)$ given
in Equation \ref{eq:L_ph}. Conservation of photon number at $r> r_{\rm
  sup}$ implies that the comoving number density evolves according to
$n_\gamma(r>r_{\rm sup}) = n_{\gamma}(r_{\rm sup}) (r_{\rm sup}/r)^2
[\Gamma(r_{\rm sup})/\Gamma(r)].$ For $n_{\gamma} \gg n_e'(r)$, one
therefore obtains $T'(r) \propto L_{th}(r)/\Gamma(r) \propto r^0$,
namely $T'(r>r_{\rm sup}) = T'_{\rm sup}$.

The electrons are continuously heated by the magnetic reconnection
process above $r_{\rm sup}$ (see Equation \ref{eq:ddotE}).  They
simultaneously radiate their energy by synchrotron emission and
inverse-Compton scattering the quasi-thermal photons. As long as the
cooling rate is sufficiently high, efficient energy transfer between
electrons and photons exit, and both populations can be characterized
by (quasi-) thermal distributions with similar temperatures, $T'_{\rm
  sup}$. However, as the jet expands, the cooling rate decreases, and
as a result, above some radius, $r_c$ ($r_{\rm sup} < r_c < r_{ph}$, see
below) the cooling can not balance the heating. At this stage,a 'two
temperature plasma' is formed, with $T'_{el} > T'_{ph} \sim T'_{\rm sup}$
\citep[see detailed discussion in][]{PMR05, PMR06}.

The radius $r_c$ at which radiative cooling balances heating is
calculated as follows.  The rate $\dot{E}$ of energy transfer via
magnetic reconnection to the plasma as it expands from radius $r$ to
$r+dr$ was calculated in Equation \ref{eq:ddotE}. We assume that about
half of this energy is used to heat the particles (the other half is
converted to kinetic energy), implying that the comoving energy gain
rate per unit volume is $P'_{heat} = (1/2) d \dot{E} /(c dV')$, where
$dV' = \Gamma(r) r^2 dr$.\footnote{The factor
  $4\pi$ is omitted since $d\dot{E}$ is already expressed in
  $\text{str}^{-1}$.} Assuming next that a fraction $f \leq 1$ of
this energy is used to heat the electrons (rather than protons), using
Equations \ref{eq:1} and \ref{eq:ddotE} the electrons
heating rate is given by \citep[][]{Giannios06}
\beq
P'_{\rm rec} = \frac{f L}{6 r^{3} \sigma_0^{3/2}} \sim 1.67 \times
10^{15} f_0 L_{52} r_{11}^{-3} \sigma_2^{-3/2}~
\rm{erg~cm^{-3}~s^{-1}}.
\label{eq:P_rec}
\eeq
The main radiative loss term of the electrons is Compton scattering
the thermal photons.\footnote{At all radii, the photon energy density,
  $u'_{th} > u_B$.} As the plasma is non-relativistic $(\gamma
\beta)^2 \simeq 3 k_B T'/(m_e c^2)$, and the power loss (per unit volume) at $r>
r_{\rm sup}$ is thus $P'_{IC} = 4 \sigma_T k_B T'_{sup} n'_{el}
u'_{th} / m_e c$ (where we assume that the thermal energy is dominated by the
photons at $r_c$). Equating the energy loss and the energy gain rates gives 
\beq
r_c = \left\{
\ba{ll}
2.0 \times 10^{11} L_{52}^{\frac{48}{85}} f_0^{-\frac{3}{5}} (\epsilon
\Omega)_3^{-\frac{59}{170}} \sigma_2^{-\frac{93}{68}} \text{cm,} &
(DC) \\
2.1 \times 10^{11} L_{52}^{\frac{129}{235}} f_0^{-\frac{3}{5}}
(\epsilon \Omega)_3^{\frac{117}{235}} \sigma_2^{-\frac{18}{235}} \left
(\frac{\bar{A}}{15} \right )^{-\frac{42}{235}} \text{cm,} & (Brem.) \\
1.5 \times 10^{11} L_{52}^{\frac{1347}{2300}} f_0^{-\frac{3}{5}}
(\epsilon \Omega)_3^{-\frac{194}{575}} \sigma_2^{-\frac{867}{575}}
\text{cm}, & (CS)
\ea
\right.
\eeq
when considering double Compton, Bremsstrahlung and cyclo-synchrotron,
respectively as the main photon production processes. We can thus
conclude that for the fiducial values of the free model parameters
chosen, $r_c < r_{ph}$ in all scenarios considered.

At radii $r_c \leq r \leq r_{ph}$, the electrons can no longer
efficiently convert their gained energy to the photons\footnote{It can
  easily be checked by integrating from $r_c$ to infinity that the
  Compton $Y$ parameter in this regime is in the order of the unity
  \citep[e.g.,][]{BSV13}.}. The photon temperature thus
freezes \footnote{The photon temperature slightly increases below the
  photosphere, due to Compton scattering with the electrons. However,
  this effect is discarded in our computation, since it only increases
  the observed temperature.}.  The peak of the observed spectrum ($\nu
F_{\nu}$) can therefore be estimated as follow.  First, as the photons
conserve their Wien distribution, the (comoving) peak energy is
slightly above the average photon temperature, $E_{p}^{'} = (4/3)
\langle \epsilon^{'} \rangle = 4 k_B T'_{\rm sup}$. Second, due to the
Lorentz boost, the observed energy of the photons that decouple the
plasma at the photosphere is $E_{pk}^{ob} \simeq 2 \Gamma(r_{ph}) E'_p$
(for on-axis observer). The peak of the observed $\nu F_{\nu}$
spectrum is therefore expected at $E_p^{ob} = 8 \Gamma(r_{ph}) k_B
T'_{\rm sup}$, namely
\begin{align}
E_{pk}^{ob}  = \left\{
\ba{ll}
 13.1\, L_{52}^{-\frac{1}{17}} (\epsilon \Omega)_3^{\frac{43}{102}}
\sigma_2^{\frac{49}{68}} \text{~~ Mev}, & (DC) \\
12.9\, L_{52}^{-\frac{4}{47}} \sigma_2^{\frac{41}{47}} (\epsilon
\Omega)_3^{\frac{70}{141}} \left (\frac{\bar{A}}{15} \right
)^{-\frac{14}{47}} \text{MeV,} & (Brem.) \\
8.1\, L_{52}^{-\frac{11}{460}} (\epsilon \Omega)_3^{\frac{151}{345}}
\sigma_2^{\frac{56}{115}} \text{~~ Mev}. & (CS).
\ea
\right.
\label{eq:E_pk}
\end{align}
Note the weak dependence on the luminosity $L$ and on $\epsilon
\Omega$, and the moderate dependence on the magnetization,
$\sigma$. 

\begin{figure}
\centering
\includegraphics[width=0.9\textwidth]{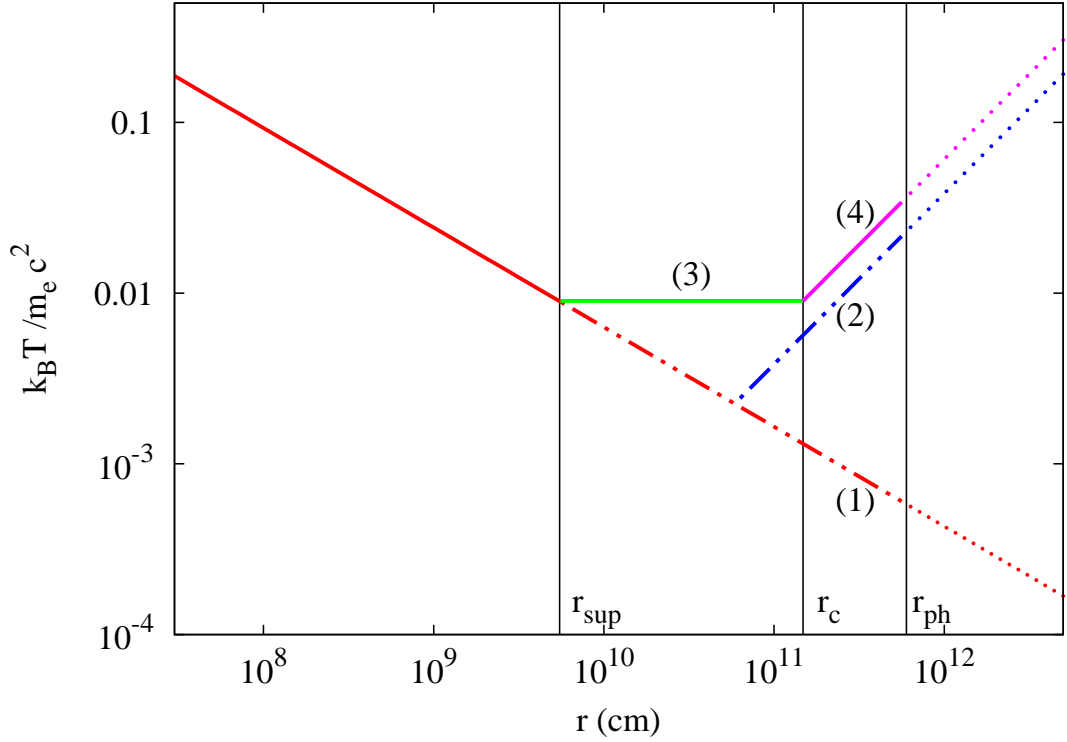}
\caption[Radial evolution of the co-moving electrons temperature for
  Poynting dominated outflows]{Radial evolution of the co-moving
  electrons temperature for the fiducial parameters $L_{52}=1$,
  $\sigma_{0,2} = 1$ and $(\epsilon \Omega)_3=1$. Four segments are
  shown. (1) At small radii, the temperature is directly associated to
  the internal energy, and $T' \propto r^{-7/12}$ (see
  Equation \ref{eq:theta}). (2) At $r > r_c$, coupling between the
  electrons and photons is weak resulting in inefficient electron
  cooling which, in turn, leads to increase of the electrons
  temperature. (3) When photon starvation is considered, the
  electron's temperature is fixed at $r_{\rm sup} < r < r_c$, $T' =
  T'_{\rm sup}$.  (4) Same as in segment (2), but when photon
  starvation is taken into account. Thus, we expect the temperature to
  evolve along segments (1) - (3) - (4) (solid lines), with
  dash-dotted lines (including segment (2)) for comparison only. With
  the parameters at hand, $r_{\rm sup}$ is given by cyclo-synchrotron
  process. }
\label{Fig:tempmagnetic}
\end{figure}

\begin{figure}
\centering
\includegraphics[width=0.9\textwidth]{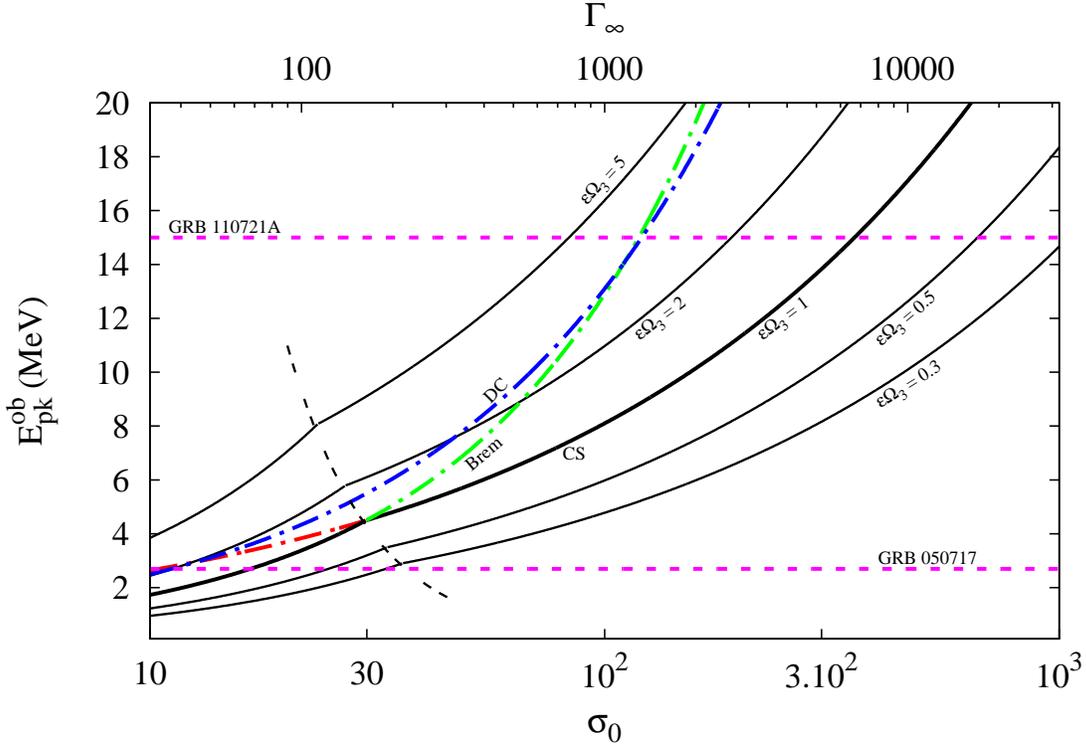}
\caption{Observed peak energy as a function of the magnetization
  $\sigma_0$ for $L_{52}=1$ and different reconnection frequencies
  $(\epsilon \Omega)_3$. Fiducial value $(\epsilon \Omega)_3 = 1$ is
  shown by the thick solid line, and other solid lines are for
  different values of $(\epsilon \Omega)_3$, as indicated. The double
  dashed thin (black) line shows the separation between photon
  production domination by bremsstrahlung process at low $\sigma_0$
  and cyclo-synchrotron domination at higher values of $\sigma_0$.
  The expected peak energy for $(\epsilon \Omega)_3=1$ via the
  different processes: bremsstrahlung, double Compton and
  cyclo-synchrotron are represented by the thick dash-dotted lines
  (green, blue and red, respectively). For comparison, we add the
  observed peak energies of the two highest record GRBs, GRB~050717
  and GRB~110721A, by the two horizontal dashed (purple) lines.}
\label{fig:Tob}
\end{figure}

The electron and photon temperatures are presented in Figures
\ref{Fig:tempmagnetic} and \ref{fig:Tob}.  In Figure
\ref{Fig:tempmagnetic} we present the spatial evolution of the
comoving electron temperature. At $r< r_{\rm sup}$, the electrons
temperature decays, in accordance to Equation \ref{eq:theta}.  At
larger radii, $r_{\rm sup} < r < r_c$, photon starvation leads to
constant temperature, and at even larger radii it increases. For
comparison, we show the results obtained when photon starvation is
omitted, which is shown by the segments (1) and (2) [dash-dotted
  lines] in Figure \ref{Fig:tempmagnetic}. In this case, the
temperature continues to decay at radii $r > r_{\rm sup}$, in
accordance with Equation \ref{eq:theta} (segment (1)). At larger radii
the temperature increases again (segment (2)), once the electrons
cannot convert efficiently their gained energy to the photons.

The dependence of the observer (photon) temperature $E_{pk}^{ob}$ on
the unknown magnetization parameter $\sigma$ is displayed in Figure
\ref{fig:Tob}. As is shown, for any value of $\sigma > 10$, the
observed peak energy is greater than a few MeV, comparable only with
the highest GRB peak energies observed.  For comparison, we provide
two examples: GRB~050717 having $E_{pk}^{ob} \sim 2.7$~MeV
\citep{KHP06} requires $\sigma_0$ to be at most in the order of 20. On
the other hand the extreme peak energy of the first seconds of GRB
110721A around $15$~MeV \citep{Axelsson+12} can be explained in a
highly magnetized jet, having $\sigma_0 \sim 350$ for $(\epsilon
\Omega)_3=1$. We stress though, that these results show that the vast
majority of GRBs, having peak energy at $<$~MeV, are inconsistent with
having high magnetization parameter, $\sigma > $~a few, at least below
the photosphere.

Within the framework of our model, a lower limit on the luminosity of
the photosphere is derived by considering the thermal luminosity at
$r_{c}$, as Compton scattering of photons above $r_c$ only increases
the photospheric luminosity. One obtains
\begin{align}
L_{th}^{ob} > \left\{
\begin{aligned}
& 4.0 \times 10^{50} L_{52}^{\frac{101}{85}} (\epsilon
  \Omega)_3^{\frac{37}{170}} \sigma_2^{-\frac{99}{68}}
  f_0^{-\frac{1}{5}} \text{~ ~ erg~s}^{-1}, & (DC) \\
& 4.0 \times 10^{50} L_{52}^{\frac{278}{235}} (\epsilon
  \Omega)_3^{\frac{352}{705}}\sigma_2^{-\frac{241}{235}}
  f_0^{-\frac{1}{5}} \left (\frac{\bar{A}}{15} \right
  )^{-\frac{14}{235}} \text{~ ~ erg~s}^{-1}, & (Brem.)\\
& 3.6 \times 10^{50} L_{52}^{\frac{2749}{2300}} (\epsilon
  \Omega)_3^{\frac{127}{575}} \sigma_2^{-\frac{864}{575}}
  f_0^{-\frac{1}{5}}\text{~ ~ erg~s}^{-1}.  & (CS)
\end{aligned} 
\right.
\label{Eq:Lthob}
\end{align}
Thus, we conclude that the thermal part of the spectrum should be at
least a few \% of the total burst luminosity. In fact, since the
non-thermal part is spectrally broad, it is possible that if observing
over a limited band, that the thermal component will carry a larger
fraction of the observed luminosity than presented here.  Such a
component, although weak, may be detected by careful
analysis. Finally, note that for GRB~110721A, the expected fraction
thermal luminosity is expected to be very small, in the order of 0.5
percent of the total luminosity.

\begin{figure}
\centering
\includegraphics[width=0.9\textwidth]{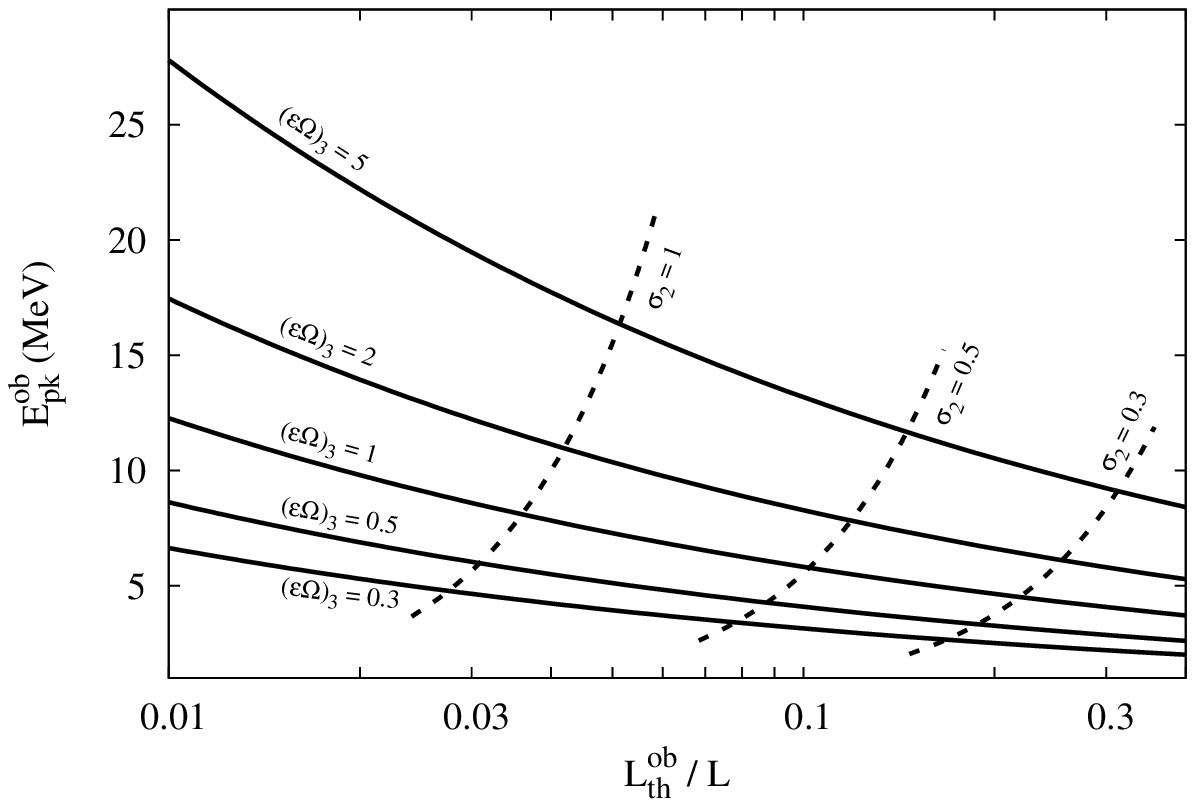}
\caption{The peak energy $E_{pk}^{ob}$ is plotted against the ratio of
  thermal to non-thermal luminosity, ($L_{th}^{ob}/L$) for different
  values of $\sigma_2$ and of $(\epsilon \Omega)$. The solid lines
  represent different values of $\sigma$ for fixed values of
  $(\epsilon \Omega)_3$, ranging from 0.3 to 5. The dashed lines
  represent different values of $(\epsilon \Omega)_3$ for constant
  $\sigma_2$, ranging from 0.3 to 1.  Here we consider $L_{52}=1$,
  however as both $E_{pk}^{ob}$ and $L_{th}^{ob}/L$ are weakly
  dependent on $L$, the result is not substantially different for
  other values of $L_{52}$.}
\label{fig:EpL}
\end{figure}

The relation between $L_{th}^{ob}/L$ and $E_{pk}^{ob}$ as a function
of $\sigma_0$ at constant $(\epsilon \Omega)$ is shown in Figure \ref{fig:EpL}.
The higher the photospheric peak (corresponding to large $\sigma_0$),
the smaller the radiative efficiency of the photosphere.

\section{Discussion}

In this work, we analyzed the expected photospheric signal from
Poynting flux dominated outflows.  As we show here, in these conditions
full thermalization can only be achieved at small radii, $r<r_{\rm
  sup} \ll r_{ph}$. As a result of this photon starvation, the
observed $\nu F_\nu$ peak of the photospheric emission is expected
above $8 \text{~MeV}$ (see Equation \ref{eq:E_pk}). Moreover, we find
that this value has only weak dependence on the unknown values of the
outflow parameters.  This value is inconsistent with the observed peak
energy of the vast majority of GRBs, which is of the order of
$350$~keV in average \citep{Ghirlanda+09, Goldstein+12} for time
integrated spectra and rises up to 3~MeV in some exceptional
cases. Thus, if this peak is due to emission from the photosphere, 
our results indicate that only the bursts with the highest
$\nu F_{\nu}$ peak might be marginally consistent with the
photospheric emission of magnetically dominated outflows.

Understanding of GRB prompt emission has been revolutionized in the
past few years, with evidence for thermal emission being widely
accepted in many bursts \citep{RP09, Ryde+10, Axelsson+12, Guiriec+11,
  Guiriec+13, Iyyani+13}. Still, a full understanding of the origin of
the spectra and the outflow conditions (in particular, the
magnetization) are far from being understood. One hint may be the
$E_{pk}- E_{iso}$ correlation (the ``Amati'' correlation;
\citet{Amati+02,Amati06}) that shows that high spectral peak energy
correlate with high total energy release. Moreover, there are some
evidence for high efficiency in the prompt emission in very energetic
bursts \citep{LZ04, Peer+12}. It was proposed that these results may
indicate a photospheric origin of a substantial part of the
observed spectra, including the peak itself \citep{TMR07, Lazzati+13,
  DZ14}.  The results obtained here show, however, that if the flows
are highly magnetized, the expected peak energy is too high to be
consistent with the observed one, and the efficiency of photospheric
emission is only a few percent (Equation \ref{Eq:Lthob}).

Our results indicate a high energy peak, at $\gtrsim 8$~MeV,
significantly higher than considered by \citet{Giannios06,
  Giannios12}. This difference originates from the larger comoving
temperature at $r_c$ obtained here, resulting from photon
starvation. They are aligned with the results obtained by
\citet{Beloborodov13}, which were considerably less detailed, and were
obtained under the assumption of initially similar thermal and kinetic
luminosity. It thus implies that the ratio of the photon number
density to the electron number density is
over-estimated. \footnote{\citet{Beloborodov13} estimated $E_{pk}^{ob}
  \simeq 3$~MeV, somewhat less than the results derived here. The
  origin of this discrepancy is his assumption of coasting
  Lorentz factor below the photosphere.}

Alternatively, the photospheric emission may be sub-dominant, the
dominant part of the prompt spectrum being non-thermal. However, in
this case, the expected peak, at $\gsim8$~MeV should contain a few
percent of the burst luminosity, and should therefore be observed.
Moreover, \citet{BP14} studied jets in which the ratio of the Poynting
luminosity to the total luminosity is large at the dissipation
zone. By identifying the MeV peak with synchrotron emission, they
found strong constraints on the dissipation radius and the Lorentz
factor at the emission region. We thus conclude that magnetized jet
models in which the photospheric component is sub-dominant have
additional difficulties with explaining the observed spectra.

High magnetization implies a lower ratio of $L_{th}/L$ (see Equation
\ref{Eq:Lthob}). Thus, the non-detection of a thermal component may be a
signal of highly magnetized outflow. Our results are therefore
consistent with the previous analysis carried by \citet{ZP09}, and
stress the need for a careful spectral analysis that could enable to
constrain the magnetization of GRB outflows.

The conditions for thermalization of the plasma in the classical
``fireball'' (when Poynting flux is sub-dominated) were studied by
\citet{VLP13}. In this work, it was shown that in order to obtain full
thermalization, the energy dissipation radius is limited to a
relatively narrow range ($r \sim 10^{10}- 10^{11}$~cm), and the
Lorentz factor during the dissipation must be mild, $\Gamma \sim
10$. Interestingly, these results are aligned with the inferred values
of the outflow parameters \citep{Peer+07}.

In the magnetized outflow scenario considered here, the dissipation
results from magnetic reconnection, and is assumed to be continuous
along the jet. Thus, one cannot constrain a particular dissipation
radius. The approach taken here is therefore different: by prescribing
the dynamics, we study its observational consequences, in particular
the expected peak energy and efficiency of the photospheric emission.

\acknowledgments DB is supported by the Erasmus Mundus Joint Doctorate Program
by Grant Number 2011-1640 from the EACEA of the European Commission. AP acknowledges support by Marie Curie grant FP7-PEOPLE-2013-CIG \#618499 

\bibliographystyle{/Users/apeer/Documents/Bib/apj}
\bibliography{/Users/apeer/Documents/Bib/abbrevs,/Users/apeer/Documents/Bib/short_abbrevs,/Users/apeer/Documents/Bib/bib_apeer}

\end{document}